\documentclass{pasj00}
\draft

\begin{document}
\SetRunningHead{Author(s) in page-head}{Running Head}
\Received{2007/04/11}
\Accepted{2007/00/00}

\title{AKARI Far-Infrared Source Counts in the Lockman Hole}


\author{
   Shuji \textsc{Matsuura}\altaffilmark{1}
   Mai \textsc{Shirahata}\altaffilmark{1}
   Mitsunobu \textsc{Kawada}\altaffilmark{2}
   Yasuo \textsc{Doi}\altaffilmark{3}
   Takao \textsc{Nakagawa}\altaffilmark{1}
   Hiroshi \textsc{Shibai}\altaffilmark{2}
   Chris P. \textsc{Pearson}\altaffilmark{1}
   Toshinobu \textsc{Takagi}\altaffilmark{1}
   Woong-Seob \textsc{Jeong}\altaffilmark{1}
   Shinki \textsc{Oyabu}\altaffilmark{1}
   and
   Hideo \textsc{Matsuhara}\altaffilmark{1}}
 \altaffiltext{1}{Department of Infrared Astrophysics, Institute of Space and Astronautical Science (ISAS), Japan Aerospace Exploration Agency (JAXA) 3-1-1, Yoshinodai, Sagamihara, Kanagawa 229-8510, Japan}
 \email{matsuura@ir.isas.jaxa.jp}
 \altaffiltext{2}{Graduate School of Sciences, Nagoya University, Furo-cho, Chikusa-ku, Nagoya 464-8602, Japan}
 \altaffiltext{3}{Department of General System Studies, Graduate School of Arts and Sciences, The University of Tokyo, 3-8-1 Komaba, Meguro-ku, Tokyo 153-8902, Japan}


%


\maketitle

\begin{abstract}
We report initial results of far-infrared observations of the Lockman hole 
with Far-Infrared Surveyor (FIS) onboard the AKARI infrared satellite. 
On the basis of slow scan observations of a 0.6 deg~$\times$~1.2 deg contiguous area, 
we obtained source number counts at 65, 90 and 140 $\rm{\mu m}$ down to 
77, 26 and 194 mJy (3$\sigma$), respectively. The counts at 65 and 140 $\rm{\mu m}$ 
show good agreement with the Spitzer results. However, our 90 $\rm{\mu m}$ counts 
are clearly lower than the predicted counts by recent evolutionary models 
that fit the Spitzer counts in all the MIPS bands. Our 90 $\rm{\mu m}$ counts 
above 26 mJy account for about 7\% of the cosmic background. These results provide 
strong constraints on the evolutionary scenario and suggest that the current 
models may require modifications.
\end{abstract}

\section{Introduction}


One of the main scientific objectives of AKARI (Murakami et al. 2007) 
is to investigate the history of galaxy evolution by measuring thermal emissions 
from interstellar dust heated by the UV light from stars. Far-Infrared Surveyor 
(FIS) onboard AKARI with four photometric bands of 65, 90, 140 and 160 $\rm{\mu m}$ 
(Kawada et al. 2007) is designed to detect such dusty objects at wavelengths near 
the peak of the dust emission. Mapping observations with AKARI in the slow scan mode 
provide 1--2 orders of magnitude higher sensitivity than that in the all sky survey mode 
with fast scan, and they are suitable for probing distant luminous infrared galaxies.

The IRAS mission has discovered the presence of luminous infrared galaxies 
in local universe (Neugebauer et al. 1984), and their number counts show 
an excess over the predicted counts for the non-evolution scenario 
(Hacking \& Soifer 1991). Far-infrared deep surveys with Infrared Space Observatory
(ISO) have confirmed the findings by IRAS at much deeper flux levels; 
faint galaxy counts at 90 and 170 $\rm{\mu m}$ show steep increase 
of the excess counts as flux levels become fainter 
(Kawara et al. 1998, Puget et al. 1999, Efstathiou et al. 2000, 
Linden-Vornle et al. 2000, Matsuhara et al. 2000, Dole et al. 2001, 
Rodighiero et al. 2003, Kawara et al. 2004).
These ISO results have inspired us with insight for the link between local 
luminous infrared galaxies and galaxy formation in early epoch, and they 
have provided useful constraints on the galaxy evolution models. 
The Spitzer Space Telescope (Werner et al. 2004) with the Multiband Imaging 
Photometer for Spitzer (MIPS) (Rieke et al. 2004) provided the number counts 
at 24, 70 and 160 $\rm{\mu m}$ down to 1--2 orders of magnitude deeper fluxes 
than previously reached and strongly constrained the galaxy evolution models 
(Papovich et al. 2004, Dole et al. 2004, Frayer et al. 2006a). The Spitzer results 
in all MIPS bands are well fitted with phenomenological models that predict most 
of the faint galaxies lie at redshifts between 0.7 and 0.9 (Dole et al. 2003, 
Lagache et al. 2003, 2004). Their models reproduce a broad shape of the measured 
redshift distribution for the Spitzer 24 $\rm{\mu m}$ samples, but the models still 
show discrepancies in some details at high redshifts $z >1.5$ 
(Perez-Gonzalez et al. 2005, Le Floc'h et al. 2005, Caputi et al. 2006). 
They suggest that some details of the discrepancies in the redshift distribution 
between observations and models may come from the field variance. 
Therefore, further study of galaxy modeling with large amount of galaxy samples 
is still required. AKARI will provide us important constraints because of its 
high statistics on precedented sensitivity levels over the whole sky. This first 
deep survey validates the AKARI great capabilities, and furthermore bring new 
constraints especially at 90 $\rm{\mu m}$, where IRAS and ISO data at faint flux 
levels below 100 mJy are not accurate enough to provide tight constraints 
on the evolutionary models.

Recent advances in observational study in wide wavelength range 
from optical to millimeter-wave have shown that luminous infrared galaxies 
are minority in local universe but major building blocks of 
the high redshift universe in terms of the total radiation energy release 
(Blain et al. 2002). Unresolved faint galaxies at high redshifts 
would form the Cosmic Infrared Background (CIB). In fact, 
CIB in the far-infrared regime measured with COBE (COsmic Background Explorer) 
accounts about half of the total energy of optical/infrared background 
(Puget et al. 1996, Hauser et al. 1998, Lagache et al. 1999, Hauser \& Dwek 2001, 
Finkbeiner et al. 2002). That tells us the importance of exploration of 
far-infrared sources which are responsible for the energy release 
in the cosmic history. Recent Spitzer results of ultra-deep galaxy counts 
at 70 $\rm{\mu m}$ (Frayer et al. 2006b) and of the stacking analysis 
for faint sources at 70 and 160 $\rm{\mu m}$ accompanied with the 24 $\rm{\mu m}$ 
sources (Dole et al. 2006) show that more than $\sim$70\% of CIB has been 
resolved into point sources. However, previous galaxy counts at the peak 
of the measured CIB ($\sim$100 $\rm{\mu m}$) in between the MIPS bands 
have not reached to such deep levels.

In this work, we performed far-infrared deep surveys with AKARI 
in the Lockman hole, in order to measure the source counts at 
previously unexplored sensitivity near the photometric bands covering 
the peak wavelength of CIB. This survey has been done in 
the performance verification (PV) phase of AKARI mission to demonstrate 
and evaluate performance of the slow scan observation. 
The survey area was, therefore, limited to 0.7 $\rm{deg^2}$, 
and east patch of the Lockman hole (LHEX field), where many observations 
including follow-up observations from the ground for the ISO survey 
(Oyabu et al. 2005, Rodighiero et al. 2005) have been done, was selected 
for the observation field so that the cross calibration between missions 
could be done. In this paper, we present the initial results of 
the number counts for the sources detected with AKARI.

\section{Observations}

The observations of the Lockman hole were carried out 
in the PV phase in 2006 May, to check the performance of 
the FIS instrument for observing faint objects and to demonstrate 
the slow-scan observations for wide-area mapping. 
The survey covers a 0.6 deg~$\times$~1.2 deg (0.7 $\rm{deg^2}$) region centered at 
RA(J2000) $=$ \timeform{10h52m12s} and DEC(J2000) $=$ +\timeform{57D18'00''}, 
the east side of the lowest cirrus region corresponding to the ISO deep survey. 

The data were taken using the FIS-02 slow-scan observational mode 
of the astronomical observation template (AOT) with 15-arcsec/s scan rate 
and 2-s period reset (Kawada et al. 2007). The FIS-02 observation provides 
a map data with double sightings of each source in a 0.13-deg wide, 
1.25-deg long strip area by a single turn-around scan. 
The mapping observation was carried out with 11 contiguous scans; 
12 scans were actually done but one scan data were lost with trouble of 
the ground station. To secure sufficient redundancy and to improve the sensitivity, 
one third of the width of each single scan strip was ovelapped with the next 
strip for producing mosaic image. It took 2 hours of telescope time in total 
to cover the entire survey field. The data in the four photometric bands, 
centered at 65 $\rm{\mu m}$, 90 $\rm{\mu m}$, 140 $\rm{\mu m}$ and 160 $\rm{\mu m}$, 
were simultaneously taken, and the same field was surveyed in all the bands 
except for a small margin of the survey area corresponding to the sight 
difference between field-of-view (FOV) of each band.

AOT includes the calibration sequence for each observation as follows. 
The dark measurements by closing the cold shutter and responsivity check 
with an internal calibrator are carried out during maneuver to change the 
satellite attitude from the all-sky survey mode to the source targeting mode. 
After the maneuver the shutter is opened, the sky signal is monitored 
during a settling time for fine control of the satellite attitude, 
and then the slow-scan observation starts. At turning point of the 
round-trip scan, the shutter is shortly closed, and responsivity check with 
the internal calibrator is done. The total exposure time for the round-trip 
scan is 10 min. The same calibration sequence as above is repeated during 
the maneuver for returning to the all-sky survey mode. In addition to 
the calibration sequence by continuous light illumination, stimulator flashes 
every 1 min to check the responsivity drift is operated as is done 
in the all-sky survey mode. Such highly redundant calibration data set 
enabled us to correct the responsivity change refering the astronomical 
calibration data taken by separate observations with the same 
calibration sequence.

\section{Data reduction and calibration}

\subsection{Initial reduction in time domain}

The raw data were initially reduced using a part of the FIS All-sky survey 
pipeline tool for the ADU to volt conversion, flagging of bad data 
(cosmic-ray events and other discontinuities) and the correction of 
non-linear integration ramps. The reduced data were processed with 
an official data analysis tool specified for the FIS slow-scan observation 
(SS-tool) to produce the basic calibrated data products and the final co-added map.

Initial process of SS-tool is the slope calculation of the ramps 
removing the cosmic-ray events identified in each ramp and 
correcting the after effect of the calibration lamp (stimuator) flashes. 
The reset interval of the integration ramp for the Lockman hole observations 
is set to be 2 seconds. The 2-s ramps are comprised of 52 and 35 samples per 
pixel for 65/90 $\rm{\mu m}$ and 140/160 $\rm{\mu m}$ detectors, respectively. 
The slope calculation processing reduces the sampling rate to 
one data per ramp per array pixel to maximize signal-to-noise and 
to avoid a periodic structure arising from incompleteness of 
the non-linear ramp correction. For 15-arcsec/s scan, the reduced data 
provides only one data sample per pixel for a source crossing 
time, but the Nyquist sampling condition is resultantly satisfied in real space 
domain after co-addition of the data using all array pixels.

Glitches and subsequent tails induced by cosmic-ray hits affect the data severely 
because of slow transient behavior of the Ge:Ga detectors. 
The integration ramps affected by the glitches were flagged out, and the tails 
with a 20 s or a longer time constant were suppressed by median filtering 
in the background subtraction processing as described later. 
The tail with a short time constant is not filtered out in this stage, 
but the affected data are removed by the sigma clipping in the co-addition process. 
The stimlator flashes also causes after effects similar to the tails by cosmic-ray hits, 
but they were repeatable in amplitude and successfully corrected using a template 
of their average time profile with an accuracy limited by the random noise.

\subsection{Flat fielding}

The second step of the SS-tool processing is to produce basic calibrated data 
for each array pixel, i.e., the flat fielding. This process is based on 
the observations of known diffuse sources; zodiacal emissions and interstellar 
dust emissions, which are expected to be smooth within the field of view (FOV). 
The diffuse-source observations were used only for the flat fielding, 
and the absolute flux calibration for point-sources was separately done 
(see Section 4), because it might be different from that for diffuse-source 
due to complex aperture effects.

Zodiacal emissions are expected to be an almost perfect flat source; 
their anisotropy in arcmin scales are smaller than 1\% (Abraham et al. 1998). 
At 65 and 90 $\rm{\mu m}$, the sky brightness of the Lockman hole is 
dominated by the zodiacal emissions, and the contributions of Galactic 
cirrus (interstellar dust) emissions, bright point sources and CIB are expected 
to be negligible. Hence, observed sky itself could be used for the flat 
fielding. The responsivity distribution of the detector arrays is derived 
from the average of time series data during the slow scan, where the average 
was calculated after removing the data that exceeds the 3-sigma noise level 
for each pixel. The flat was built from the data during the first half of the 
single turn-around scan of each pointed observation, because the latter part 
of the data was affected by the stray light as described in the following section. 
The flat fielding is done by dividing the data by the responsivity distribution. 
With this "self" flat fielding method, any stripes in the image caused by 
the flat field error were buried under the random noise.

At 140 and 160 $\rm{\mu m}$, detector signals are dominated by 'offset' light 
with a constant intensity from the internal light source. The offset light 
is implemented to improve response speed of the stressed Ge:Ga detector 
specified for fast scan of the all sky survey mode (more than 10 times 
faster than the slow scan used for the Lockman hole survey). 
Although Galactic cirrus emissions at high latitudes could be a flat source 
with moderate smoothness ($\sim$10\%) and relatively high brightness, 
their signals are less than 10\% of that of the offset light. Intensity 
distribution of the offset light at the aperture of detector arrays is 
estimated from the laboratory measurements to be uniform within 10\%. 
Sky brightness towards the Lockman hole at 140 and 160 $\rm{\mu m}$ 
is less than $\sim$2\% of the offset light, and its fluctuation is also 
negligible. Hence, the self flat fielding method with the average sky signal 
including the offset light was applied to correct the responsivity variation 
in the detector arrays, as was done for the 65 and 90 $\rm{\mu m}$ bands.

\subsection{Subtraction of sky background}

In order to emboss faint point-sources on the co-added map, 
subtraction of the sky background is required unless its brightness 
is constant in the field. Unexpectedly, there exists a stray light of 
earth limb emission onto the focal plane under a condition of small 
earth avoidance angles of the telescope, which varies with attitude 
and orbit of the satellite. The stray light shows not only a long-term 
variation depending on seasonal change of the orbital inclination 
but also short-term variation during a single observation due to change 
of the earth avoidance angle at every moment. Brightness of the stray light 
has the maximum (2--3 MJy/sr at 90 $\rm{\mu m}$) at the beginning and 
the end of the slow scan, and the minimum and plateau region appear 
around the mid time.

To subtract such slowly varying components as the stray light, 
we applied the median filtering to the data in time domain with 
a window size of 20 data samples corresponding to 10 arcmin ($\sim$15 times FWHM). 
Then, the smoothed component was subtracted from the data. 
As a result of this filtering, extended structure larger than roughly 
a half of the window size disappears from the map. 

\subsection{Co-addition and position accuracy}

The final step of the SS-tool processing is the co-addition of 
the calibrated time series data onto a sky grid. The sky position 
of each data point was derived from the telescope boresight 
according to the satellite attitude and from the array pixel map 
on the focal plane. The grid pixel sizes for all the wave bands 
were set to 30 arcseconds to secure redundancy, i.e., sufficient number of 
data per pixel ($>$5 on the average). In the co-adding process, small glitches 
and other artifacts were sigma-clipped with the standard deviation 
calculated at each grid pixel. Threshold for the sigma-clipping was set 
to 1.5 times standard deviation. Percentage of the rejected data 
in this process was about 3\% of the original data.

Co-addition of the data that were taken in different scans requires 
accurate sky position of each array pixel. The telescope boresight 
has been calibrated with the slow-scan observations of stars and asteroids, 
and the resultant position accuracy is approximately 10 arcseconds 
in absolute measure. The array pixel map is based on the optical simulation 
and its in-flight confirmation with point-source observations. 
The image distortion has not been fully measured in the entire FOV, 
but the in-flight measurements at limited positions have been carried out 
with observations of point sources including a bright source in the field. 
As a result of the measurements, the simulation has been verified with 
an accuracy of 10 arcseconds, which is limited by the absolute position accuracy.

\subsection{Final co-added maps}

\begin{figure}
  \begin{center}
    \FigureFile(150mm,150mm){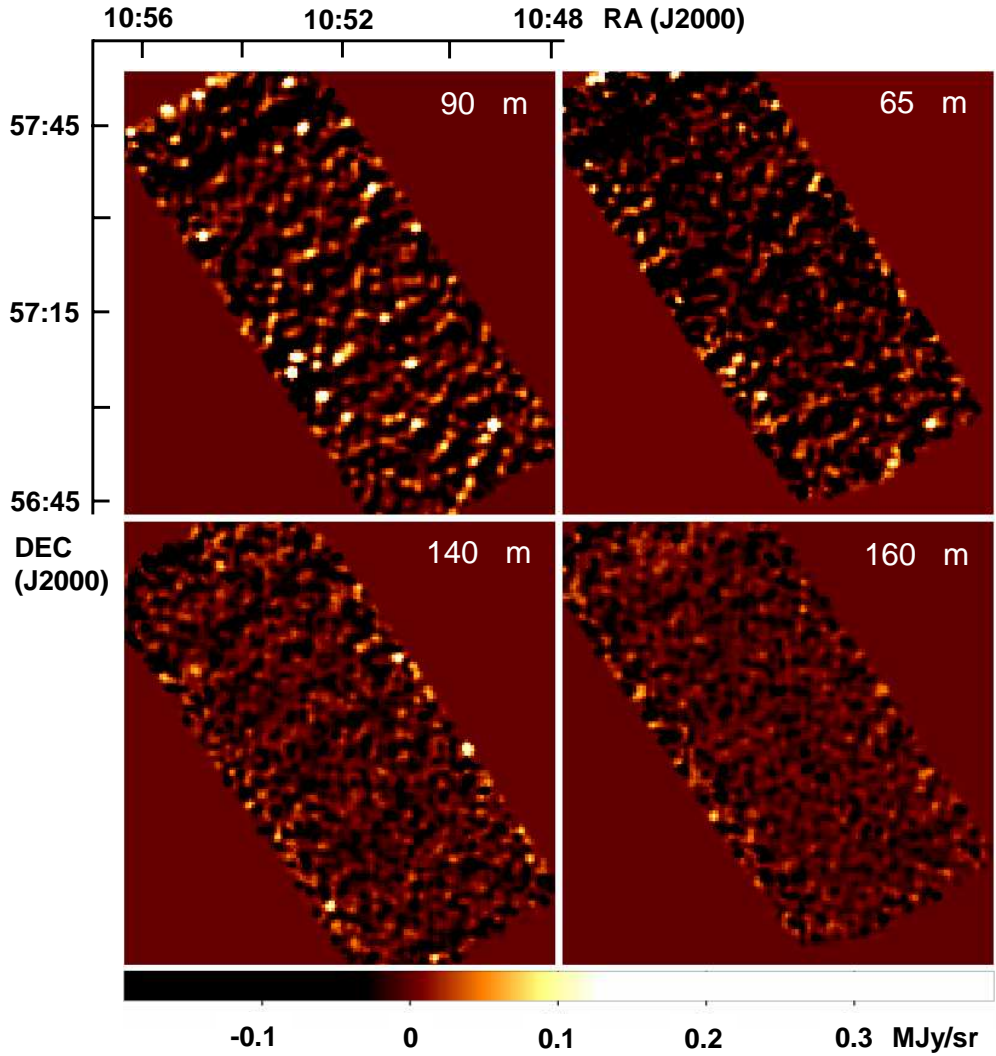}
  \end{center}
  \caption{The final co-added map of the Lockman hole field 
  in all photometric bands in equatorial coordinates with 
  the pixel size of 30 arcsec. Upper-left: 90 $\rm{\mu m}$, 
  Upper-right: 65 $\rm{\mu m}$, Lower-left: 140 $\rm{\mu m}$, 
  and Lower-right: 160 $\rm{\mu m}$. The image size is approximately 
  1.2 deg in RA $\rm{\times}$ 1.4 deg in DEC, and the observed area 
  is 0.7 square degrees. 
  Gaussian smooth filtering with a window size of 1 arcmin 
  is applied to all images.}
  \label{fig:fig1}
\end{figure}

In Fig.~\ref{fig:fig1}, we show the final co-added maps of 
the Lockman hole field at all wavelengths in equatorial coordinates 
with a pixel size of 30 arcseconds. The image after Gaussian smoothing 
with a window size of 1 arcminute is shown as linear contours 
in the unit of surface brightness. The pixel noise of each map was 
computed and summarized in the last column of Table \ref{tab:table1} 
in surface brightness. At 65, 140 and 160 $\rm{\mu m}$, the pixel noises 
are dominated by instrumental noise. At 90 $\rm{\mu m}$, the pixel noise 
is contributed by the source noise, which can be recognized as 
a positive tail exceeding the standard deviation in the histogram 
of pixel distribution. The noise difference between the 65 and 90 $\rm{\mu m}$ 
data arises from differences in optical efficiency and filter bandwidth, 
while the detectors for these two bands are identical. The noise values for 
the 140 and 160 $\rm{\mu m}$ bands are different for the same reason. 
Consequently, the 90 $\rm{\mu m}$ band is the most sensitive in all 
the photometric bands.

\section{Photometry and point-source calibration}

\subsection{Source extraction}

To extract point sources from the final co-added image, 
we used photometry tools, FIND, GCNTRD and APER, in the IDL Astronomy User's 
Library at NASA/GSFC (Landsman 1993). 
The tools search center positions of point sources with a Gaussian window 
function approximating the Point Spread Function (PSF) and measure 
the fluxes by aperture photometry. PSF in each wavelength band is derived 
from many observations in the PV phase and for calibration. The aperture 
radius was set to the full width half maximum (FWHM) of PSF. The aperture 
photometry was carried out without the sky subtraction, which was already 
done in the median filtering process. This sky subtraction method is helpful 
to avoid losing sources near a defect pixel caused by some artifacts.

\subsection{Photometric calibration}

Photometric calibration used in this paper is based on 
point-source observations in the PV phase in various fields, 
because of lack of well-calibrated source in the Lockman hole. 
According to the measured PSF, the photometric signals were compared 
with the expected fluxes of the calibration sources converted  to the 'flat' spectra, 
$\nu \times F_{\nu} =$ constant, and their linear correlation factors 
were derived. The expected fluxes of the calibration sources range from 
2 Jy to 200 Jy at 90 $\rm{\mu m}$, and deviation from the linear relation 
had no flux dependence in such a wide flux range. The calibration accuracy 
estimated from the standard deviation for various measurements 
was 17\%, 16\%, 12\% and 22\% at 65, 90, 140 and 160 $\rm{\mu m}$, respectively. 
The point-source noise in flux unit, i.e. the 1-sigma detection limit for point source, 
is derived from the pixel noise in surface brightness and the aperture correction factor, 
as summarized in Table~\ref{tab:table1}.

\subsection{Color uncertainty}

As the flux calibration for AKARI refers to the flat spectrum, 
the real flux is obtained by the color correction depending on Spectral 
Energy Distribution (SED) of the source. However, it is difficult to 
measure the infrared color for all the detected sources especially 
at faint flux levels, because most of the sources were detected 
at 90 $\rm{\mu m}$ only. Hence, for producing the source counts 
we assume that SED of all sources have the flat spectrum, and we 
estimate the uncertainties of fluxes for various SEDs.

\begin{figure}
  \begin{center}
    \FigureFile(120mm,120mm){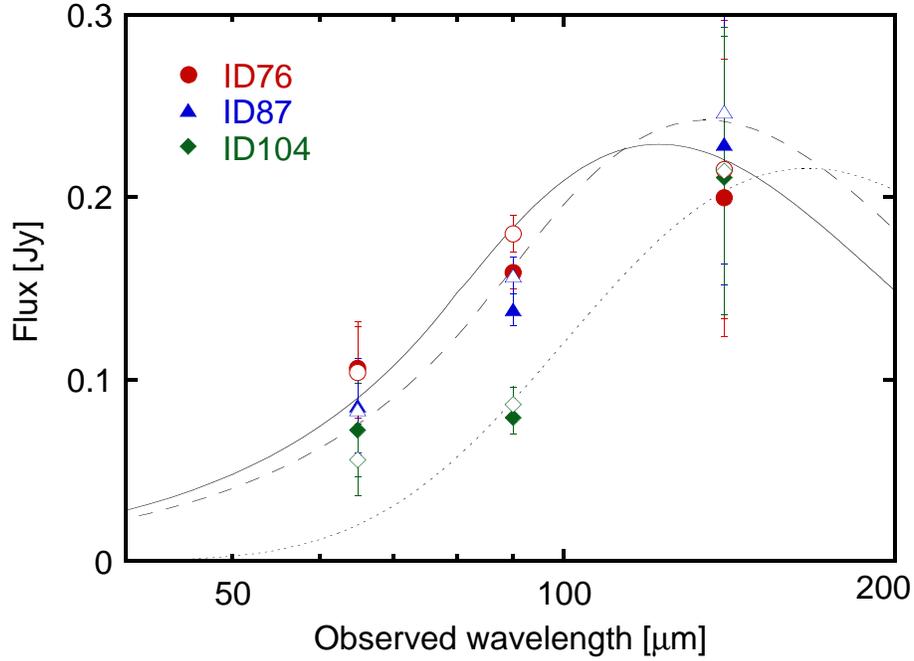}
  \end{center}
  \caption{Examples of the measured SEDs of bright sources in Lockman hole 
  (ID76: circles, ID87: triangles and ID104: diamonds) as functions of 
  the observed wavelength are compared with modified graybody spectra; 
  $T=$27 K, $\beta=$1.5 and $\alpha=$2.4 for ID76 (thin line), $T=$23 K, 
  $\beta=$1.5 and $\alpha=$2.4 for ID87 (dashed line), and $T=$25 K, $\beta=$1 
  with a redshift of $z=$ 0.362 (effectively $T=$18 K) for ID104 (dotted line). 
  Filled and open symbols denote the data without and with the color correction, 
  respectively. The model fluxes are scaled to fit the measured fluxes.}
  \label{fig:fig2}
\end{figure}

The color correction was available only for bright sources detected 
in multiple wavelength bands of AKARI. In Fig.~\ref{fig:fig2}, 
color-corrected (open symbols) and uncorrected (filled symbols) spectra 
of three bright sources in the Lockman hole are shown. The data are 
compared with modified graybody spectra with different color temperatures; 
combinations of a graybody spectrum with an emissivity $\beta$, 
$F \sim \nu^{\beta} \times B(T)$, and natural extension from the graybody 
towards shorter wavelengths by a power-law spectrum with an index $\alpha$, 
$F \sim \nu^{-\alpha}$, (e.g., Blain et al. 2002). The error bars are 
the point-source noise in Table~\ref{tab:table1}. These sources have been 
detected by the ISO survey and optically identified, and their redshifts 
have been measured by spectroscopic observations (Oyabu et al. 2005). 
Two of the sources, ID76 and ID87, are nearby star-forming galaxies at 
$z =$ 0.08 and 0.09, and the color correction is up to 10\% in all bands. 
The modified graybody models ($T=$27 K, $\beta=$1.5 and $\alpha=$2.4 for 
ID76, $T=$23 K, $\beta=$1.5 and $\alpha=$2.4 for ID87) 
give good fits to the measured spectra even for uncorrected data. 
Another source (ID104) is ULIRG with a moderate redshift of $z =$ 0.362. 
A graybody model with $T=$25 K (effectively $T=$18 K corresponding to the redshift), 
$\beta=$1 and no power-law component provides reasonable fit to the data. 
The color correction for the 65 $\rm{\mu m}$ data is 29\%, 
which is an extreme case in the parameter range of color temperature. 
If the assumed color for ID104 is correct, the real 65 $\rm{\mu m}$ flux 
is expected to be below the detection limit, but the observed flux shows 
a finite value at 65 $\rm{\mu m}$. The high flux at 65 $\rm{\mu m}$ 
may be due to noise-induced flux boosting, which is similar to Eddington bias 
or Malmquist bias and usually seen for sources near the detection limit 
(Heraudeau et al. 2004).

Frayer et al. (2006a) reported that the Spitzer sources detected at 24, 70 
and 160 $\rm{\mu m}$ in the xFLS field have a typical dust temperature of 
30 K with an emissivity of $\beta=$1.5 and a power-law index of 
$\alpha=$2.4. This color temperature is consistent with 15--25 K 
for $\beta=$2 derived from the European Large Area ISO Survey (ELAIS) 
(Heraudeau et al. 2004) and from the ISO deep survey in Lockman hole (Oyabu et al. 2005). 
According to these results, most of far-infrared sources have color temperatures 
ranging from 15--40K with $\beta=$1--2 and of the power-law indices of 
$\alpha=$1--2.5. Uncertainty of the flux measurement for the sources with colors 
in this parameter range is estimated to $+$4/$-$38\%, $+$15/$-$6\%, 
$+$9/$-$4\% and $+$2/$-$1\% at 65, 90, 140 and 160 $\rm{\mu m}$, respectively. 
In most cases for high temperature sources, the color correction uncertainties 
are smaller than the flux calibration errors. For nearby cold dust components 
or high redshift sources, the uncorrected flux at 65 $\rm{\mu m}$ 
assuming the flat spectrum could be greatly overestimated.

\section{Source counts}

\subsection{90 $\rm{\mu m}$ counts}

The final catalogs were first produced at 90 $\rm{\mu m}$, at which 
the deepest data were obtained. Sources with signal-to-noise ratios 
of S/N$>$3 were extracted according to the point-source noise shown 
in Table~\ref{tab:table1}, while a sharp decrease in number counts due to 
the incompleteness appeared at S/N$\sim$1.8 (16 mJy). At 90 $\rm{\mu m}$, 85 sources 
were found in the entire survey field down to 26 mJy. The surface density 
of sources derived from the beam size, $\Omega=4.5\times10^{-8}$ sr, 
is $\sim$70 beams per source. This is far above 22 beams per source that 
is required for the 90\% completeness without source confusion estimated 
by Helou \& Beichman (1990) and also much greater than 27 beams per source 
as a criterion for the 3-sigma detection in case of the Euclidian counts 
(Franceschini 2000).

\begin{figure}
  \begin{center}
    \FigureFile(120mm,120mm){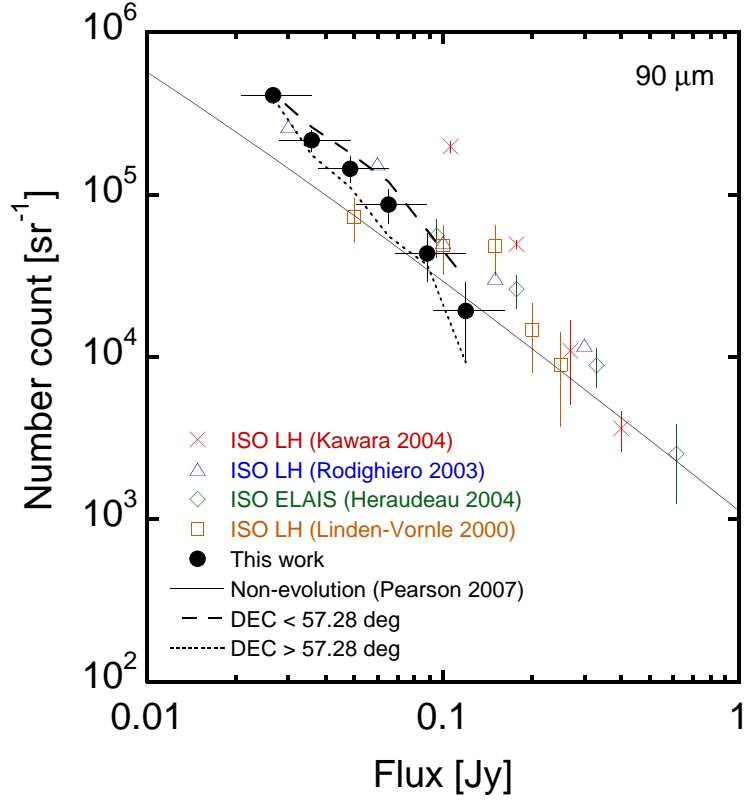}
  \end{center}
  \caption{Integral counts at 90 $\rm{\mu m}$ with no correction 
  for completeness (filled circles). Previous results of the ISO surveys 
  at 90 $\rm{\mu m}$ in the Lockman hole are shown; Linden-Vornle et al. 2000 
  (open squares), Kawara et al. 2004 (crosses), Rodighiero et al. 2004 
  (open triangles). The ISO counts in the ELAIS field (Heraudeau et al. 2004) 
  are plotted with diamonds. The solid line is a non-evolution model 
  by Pearson (2007). The dashed and dotted lines are the source counts 
  for sub-samples in the fields of DEC $<$ 57.28 deg and DEC $>$ 57.28 deg, 
  respectively (see text). }
  \label{fig:fig3}
\end{figure}

In Fig.~\ref{fig:fig3} the integral counts at 90 $\rm{\mu m}$ are plotted 
as filled circles. The data are raw counts not corrected for incompleteness. 
Error bars along the count axis are 1-sigma Poisson uncertainty, 
and the flux errors correspond to total uncertainties including 
both absolute calibration and color correction. 
Our results are compared with those from the ISO surveys at 90 $\rm{\mu m}$ 
towards the Lockman hole (Linden-Vornle et al. 2000, Kawara et al. 2004, 
Rodighiero et al. 2003) and the ELAIS field (Heraudeau et al. 2004). 
A no-evolution model by Pearson (2007) is also shown as a reference.

Our source counts in the Lockman hole reaching to $\sim$25 mJy at S/N$>$3 are 
roughly two times deeper than those of ISO surveys. Hence, 
it is not easy to compare our data with previous results at the faint end. 
At shallower flux levels our data can be directly compared with the ISO results, 
because a large part of our survey area is overlapped with the ISO observed field. 
At $\sim$100 mJy, our counts show good agreement with all the ISO results except 
for Kawara et al. (2004). A small discrepancy at the bright end could be due to 
large statistical errors of our counts. 

At fainter flux levels, our integral counts seem to agree with 
Rodighiero et al. (2003) at 60 and 30 mJy. However, individual sources 
in the counts listed in Rodighiero et al. (2003) are not always consistent 
with our measurements in both flux and position, while relative fluxes of 
all the sources listed in Kawara et al. (2004) are well aligned with our data, 
as described in Appendix. 
Our counts do not show the strong excess reported in Kawara et al. (2004) and 
also in related papers (Kawara et al. 1998, Matsuhara et al. 2000). 
This discrepancy is partly explained by different flux calibration by a factor of 
$\sim$2 (see Appendix). Simple linear scaling of their data to ours can be 
reasonably done at fluxes brighter than 100 mJy, but it fails at fainter levels 
because of steep rise of their counts. In spite that the ISO results by 
Rodighiero et al. (2003) and Kawara et al. (2004) in the Lockman hole were 
derived from the same data set, their results show large difference with each other. 
This fact suggests the difficulty in reduction of the ISO 90 $\rm{\mu m}$ data; 
they have suffered from glitches and tails induced by cosmic rays 
(Franceschini et al. 2001). 

In order to check the effect of the field variance to the counts, 
we divided our observed field into two at DEC $=$ 57.28 deg as number of the detected 
sources for the two sub-samples are equal to each other, and then we produced 
the source counts for each sub-samples. In Fig.~\ref{fig:fig3} the source counts 
in the fields of DEC $<$ 57.28 deg and DEC $>$ 57.28 deg are plotted as dashed 
and dotted lines, respectively. It is clear that at brighter flux levels the lower 
DEC field contains about twice more sources than the higher DEC field, 
while at the faint end the counts in the two fields agree with each other. 
The field variance for this case is significant exceeding the Poisson uncertainty.

In Fig.~\ref{fig:fig4} the same 90 $\rm{\mu m}$ counts are shown 
in differential form $dN/dS$ normalized to the Euclidean law $N \sim S^{-2.5}$ 
with error bars of 1-sigma Poisson uncertainty. Total uncertainty of 
absolute calibration and color correction is indicated with a cross symbol 
at lower left (at 20 mJy) of the figure. The results are summarized in 
Table~\ref{tab:table2} together with the integral count data. Our differential 
counts are compared with previous results from the same references 
as the integral counts. Again, our counts are slightly lower than that for 
the previous observations, but both the ISO results except for Kawara et al. (2004) 
and our results show a general tendency of flat counts with no steep rise or drop 
in the measured flux range.

\subsection{Counts at other wavelengths}

At wavelengths other than 90 $\rm{\mu m}$, the sources accompanied with 
the 90-$\rm{\mu m}$ sources were selectively extracted. 
In terms of real source extraction against any artifacts, 
the catalog produced in such a way is more reliable than that 
individually produced at each wavelength. The 90-$\rm{\mu m}$ selected catalogs 
are possibly biased as to miss exceptionally hot or cold/high-z populations. 
However, such biasing effects are expected to be small. Because, the detection 
limit at 90 $\rm{\mu m}$ is much better than that at the other wavelengths, 
and any detected source having ordinary temperatures of $T=$15--40K with 
$\beta=$1--2 should be identified at 90 $\rm{\mu m}$.

To extract the commonly detected sources, a centroid near a 90-$\rm{\mu m}$ 
source was searched at each wavelength allowing a small position difference 
corresponding to the beam size. Only signals associated with the 90 $\rm{\mu m}$ 
sources having $S/N >$5 were regarded as real sources. At 140 $\rm{\mu m}$, 
the signal-to-noise threshold for the source extraction was lowered to 
$S/N >$2.5 to secure statistically significant number of samples, while 
the threshold at 65 $\rm{\mu m}$ is set to $S/N >$3.
The signals rejected by this criteia showed relatively large position 
differences from the 90 $\rm{\mu m}$ sources compared with the real sources. 
As a result, 11 and 6 sources were found at 65 and 140 $\rm{\mu m}$, respectively. 
It is noteworthy that all the 6 sources at 140 $\rm{\mu m}$ were 
identified at 65 $\rm{\mu m}$. At 160 $\rm{\mu m}$, no signal 
matches to the criteria for source extraction.

In Fig.~\ref{fig:fig5} and Fig.~\ref{fig:fig6}, differential 
source counts at 65 and 140 $\rm{\mu m}$ are plotted with filled circles. 
Error bars for the counts are 1-sigma Poisson uncertainty. 
The flux error corresponding to both absolute calibration error and 
color correction uncertainty are shown by a cross symbol in each figure. 
Our results are compared with Spitzer counts at 70 and 160 $\rm{\mu m}$, respectively, 
in GTO fields (Dole et al. 2004) and xFLS field (Frayer et al. 2006a). 
The Spitzer data were converted at 65 $\rm{\mu m}$ and 140 $\rm{\mu m}$ with 
a flat spectrum of $\nu \times F_{\nu} =$ const as assumed in their paper. 
The Spitzer GTO data in the intermediate flux range show a field variance 
between those taken in two different fields (Marano and CDFS; Chandra Deep Field South). 
At both wavelengths, our counts are slightly higher than the Spitzer counts 
but agree with them within error bars and the field variance.

In Fig.~\ref{fig:fig4}, partial number counts for the 90 $\rm{\mu m}$ sources 
that are constituents of the 65 $\rm{\mu m}$ counts are plotted with an open circle. 
About half of the 90 $\rm{\mu m}$ counts in a range from 70 to 160 mJy 
consists of the 65 $\rm{\mu m}$ sources. The average color temperature of 
these sources without color correction is $T$$\sim$40 K for $\beta$=1.5, 
corresponding to the color ratio $S90/S65\sim1$. This temperature 
is hotter than that of the Spitzer sources in the xFLS field, $T$$\sim$30 K 
for $\beta$=1.5 (Frayer et al. 2006a).  The main reason that our 65 $\rm{\mu m}$ 
counts are slightly higher than the Spitzer counts is the field variance; 
most of the 65 $\rm{\mu m}$ sources lie in the lower DEC field where more bright 
sources exist, as described in section 5.1. Note that our 140 $\rm{\mu m}$ counts 
are also consistently higher than the Spitzer 160 $\rm{\mu m}$ counts. 
The remained half sources in the 90 $\rm{\mu m}$ counts would have lower color 
temperatures as $T$$<$30 K to be consistent with the Spitzer results. 
Such redder color populations may contribute to the 65 $\rm{\mu m}$ counts 
at fluxes below the detection limit.

\section{Discussions}

As described in the last section, our source counts at 65 and 140 $\rm{\mu m}$ 
are fairly consistent with Spitzer observations. In this sense, our count data 
at 90 $\rm{\mu m}$, which lie in between 65 and 140 $\rm{\mu m}$, are simply 
expected to agree with recent evolutionary models that show excellent agreement 
with the Spitzer counts. In the following, we describe the comparison of 
our counts with the models.

\begin{figure}
  \begin{center}
    \FigureFile(120mm,120mm){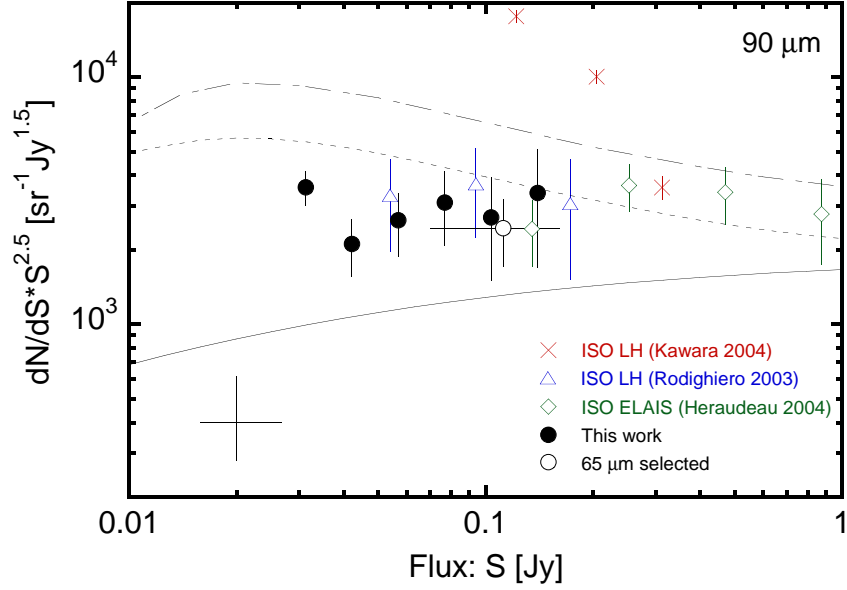}
  \end{center}
  \caption{Differential counts at 90 $\rm{\mu m}$ with no correction 
  for incompleteness. The ISO results are plotted with the same symbols 
  as Fig.~\ref{fig:fig3}. A cross symbol at lower left shows total 
  uncertainty including flux calibration error and color correction 
  uncertainty (see text). Recent evolutionary models are also shown; 
  solid and dashed lines: non-evolution and evolution model by Pearson (2007), 
  and dash-dotted line: Lagache et al. (2003) taken from Heraudeau et al. (2004). }
  \label{fig:fig4}
\end{figure}

\begin{figure}
  \begin{center}
    \FigureFile(120mm,120mm){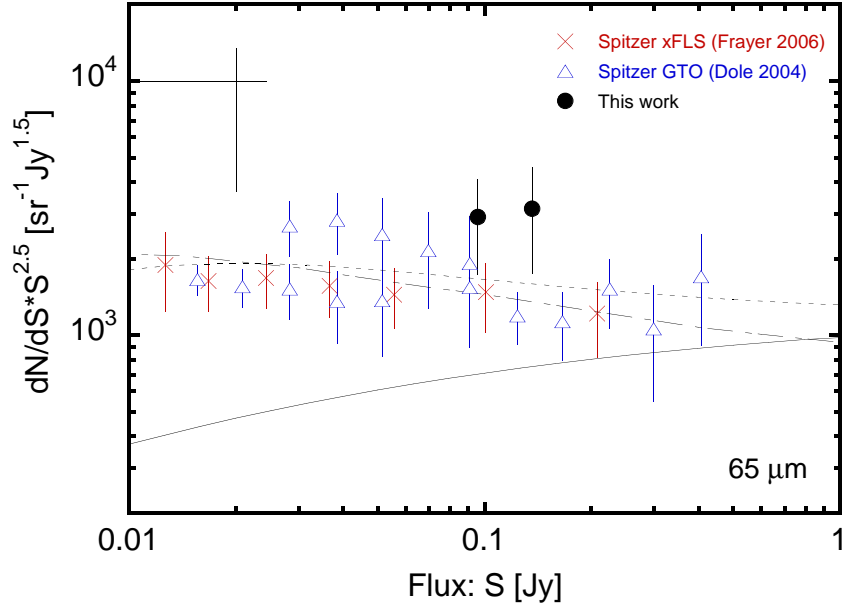}
  \end{center}
  \caption{Differential counts at 65 $\rm{\mu m}$ for the 90 $\rm{\mu m}$ 
  selected sources (filled circles) compared with the Spitzer counts at 
  70 $\rm{\mu m}$ scaled to 65 $\rm{\mu m}$ (triangles: Dole et al. 2004, 
  crosses: Frayer et al. 2006a). A cross symbol at upper left shows total 
  uncertainty of the measured flux (see text). 
  The volutionary models same as Fig.~\ref{fig:fig4} are also shown.}
  \label{fig:fig5}
\end{figure}

\begin{figure}
  \begin{center}
    \FigureFile(120mm,120mm){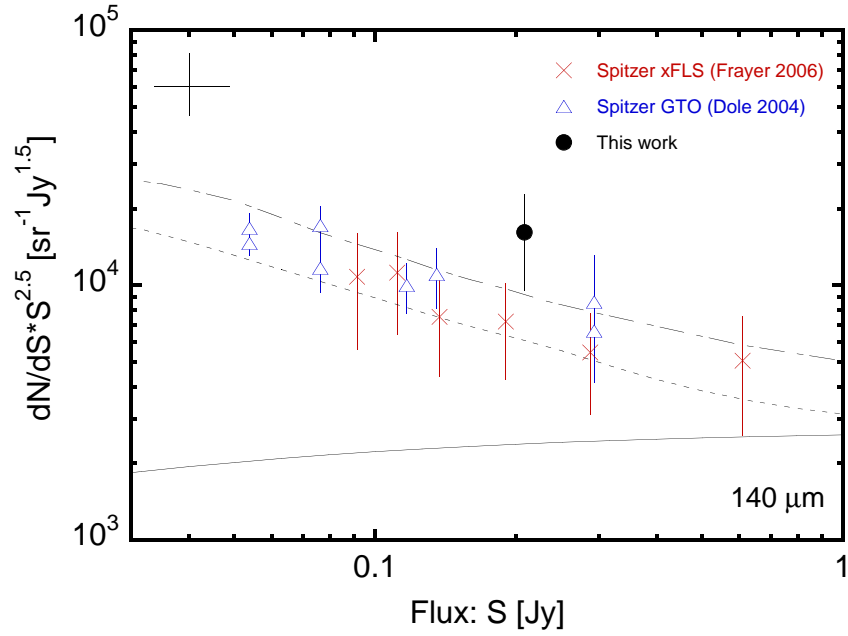}
  \end{center}
  \caption{Differential counts at 140 $\rm{\mu m}$ plotted with the same symbols 
  as Fig.~\ref{fig:fig5}.}
  \label{fig:fig6}
\end{figure}

In Fig.~\ref{fig:fig4}, \ref{fig:fig5} and \ref{fig:fig6}, 
source counts predicted from models at each wavelength are shown; 
dashed: Pearson (2007) and dash-dotted: Lagache et al. (2003, 2004). 
In Fig.~\ref{fig:fig4}, Lagache et al. (2003) model taken from Heraudeau et al. (2004) 
was used, because the model counts at the wavelength near 90 $\rm{\mu m}$ is not 
presented in Lagache et al. (2004), but the difference between their models 
of 2003 and 2004 are negligible in the AKARI wavelength bands (Lagache et al. 2004). 
For Fig.~\ref{fig:fig5} and \ref{fig:fig6}, the predicted counts at 
70 and 160 $\rm{\mu m}$ in Lagache et al. (2004) were converted to 
65 and 140 $\rm{\mu m}$, respectively, assuming the flat spectrum.
A non-evolution model by Pearson (2007) is also shown as a reference.
Lagache et al. (2003, 2004) uses phenomenological approach aiming to build 
the simplest model. They assume that infrared galaxies are mostly powered 
by starformation and use typical SEDs of normal and starburst galaxies. 
The luminosity function is represented by these two activity types. The model 
parameters are adjusted to fit the number counts in multi-wavelength bands 
from mid-infrared to sub-mm, the redshift distributions, the local luminosity 
functions at 60 and 850 $\rm{\mu m}$, and CIB.
The contemporary galaxy evolution model of Pearson (2007) uses a backward
evolution process based on the IRAS all-sky PSCz multi-component
luminosity function defined at 60 $\rm{\mu m}$ comprising of cool normal quiescent
galaxies, and a warmer component defined by infrared luminosity as
$L_{IR}<10^{11}L_{\odot}$ starburst galaxies,
$L_{IR}>10^{11}L_{\odot}$ LIRG sources, $L_{IR}>10^{12}L_{\odot}$
ULIRG sources and AGN.
Both luminosity evolution and density evolution is included in the models.
This model fits the observed source counts from the 2--1200$\rm{\mu m}$ from 
the IRAS, ISO, Spitzer missions and the SCUBA/JCMT,
MAMBO/IRAM instruments.
Differences between these models are small at brighter fluxes, 
but at flux levels below 100 mJy strong evolution effects set in 
and significant differences between models appear.

A significant result of the model comparison is that the observed counts 
at 90 $\rm{\mu m}$ in a flux range of 26--160 mJy are apparently lower than 
the predicted counts using a model by Lagache et al. (2003, 2004), which show 
excellent agreement with Spitzer counts in all MIPS bands of 24, 70 and 160 $\rm{\mu m}$ 
and also with our data at 65 and 140 $\rm{\mu m}$, as shown in Fig.~\ref{fig:fig5} 
and ~\ref{fig:fig6}. Descrepancy between the model and data is clear 
at fainter levels; the model continues to increase while the observed counts keep 
constant. The model by Pearson et al. (2007) shows better fit with our 90 $\rm{\mu m}$ 
counts, but it still predicts slightly too high counts to fit the data. 
Other models, e.g. Chary and Elbaz (2001), King and Rowan-Robinson (2003), 
and Balland, Devriendt and Silk (2003), also reproduce various observables including 
the source counts for the infrared luminous galaxies within uncertainties 
of the observation data, and all of them predict similar counts to above two models. 
Consequently, no model to explain the low 90 $\rm{\mu m}$ counts is currently available.

The incompleteness of our data might lower the 90 $\rm{\mu m}$ counts at fainter 
flux levels, but it is unlikely, because the 65 and 140 $\rm{\mu m}$ counts derived 
from the 90 $\rm{\mu m}$ selected catalogs are consistent with or even higher than 
the Spitzer data and the models. Separation between models and data starts already 
at $\sim$80 mJy, where the signal-to-noise is high ($S/N\sim10$) enough not to lose 
many sources. If the discrepancy is due to the flux calibration, our flux calibration 
has to be different by a factor of $\sim$2, which is unreasonably larger than our estimate 
of the calibration error. In order to explain our results, the evolutionary models 
may require some modifications.

It should be emphasized that our results provide strong constraints 
on the evolutionary scenarios. A new model to explain our results may be 
based on more complicated SEDs than previously used; faint galaxies 
may have exceptionally low emissivity at 90 $\rm{\mu m}$ due to dust properties. 
There also exists a room to tune model parameters to represent high redshift 
components (e.g. ULIRG), because their redshifts and SEDs are not well defined. 
In any cases, the AKARI observations of much larger number of SED samples 
are essential for constructing the new model.
Combination of source counts and redshift distributions are much stronger 
discriminators for various evolutionary models (e.g. Le Floc'h et al. 2005). 
Although the redshift measurements by optical spectroscopy for some of 
the ISO 90 $\rm{\mu m}$ sources have been done (Oyabu et al. 2005, 
Rodighiero et al. 2005), only 9 spectroscopic samples in their catalogs match 
with the AKARI 90 $\rm{\mu m}$ sources. In the spectroscopic samples, 8 sources 
have redshifts of $z < 0.4$, and the remaining 1 source is $z = 1.1$. 
Such statistically poor samples with low redshifts cannot constrain any 
evolutionary models. Further measurements and analysis of the redshift 
distribution of the AKARI 90 $\rm{\mu m}$ sources are required.

Unresolved sources below the detection limit form the CIB. The integrated flux 
of our 90 $\rm{\mu m}$ counts down to 26 mJy corresponds to the surface 
brightness of 0.021 MJy/sr. This is only $\sim$7\% of the CIB estimated 
by Lagache et al. (2003) and Dole et al. (2006). Dole et al. (2004) reported 
that the Spitzer counts in the GTO field at 70 and 160 $\rm{\mu m}$ down to 
15 and 50 mJy account for about 23\% and 7\% of the CIB, respectively. 
For the xFLS survey down to 8 and 50 mJy, approximately 35\% and 15\% 
of the CIB were resolved at 70 and 160 $\rm{\mu m}$, respectively (Frayer et al. 2006a). 
Frayer et al. (2006b) also claimed that ultradeep Spitzer counts down to 1.2 mJy 
account for about 60\% of the CIB at 70 $\rm{\mu m}$ and that a turn-over 
of the differential counts, as seen in the Spitzer 24 $\rm{\mu m}$ counts 
(Papovich et al. 2004), appears at $\sim$10 mJy. If we simply assume the flat 
spectra for all galaxies, the turn-over point of the 90 $\rm{\mu m}$ counts 
is expected to be $\sim$13 mJy. These results suggest that number of the bulk 
galaxies of the CIB lying below the point-source detection limit at 90 $\rm{\mu m}$ 
may steeply increase towards the turn-over point. In order to search for such 
missing sources, much deeper survey at 90 $\rm{\mu m}$ is demanded. The flux 
limit of the present survey is still above the confusion limit, and source 
counts to deeper levels can be obtained by spending much exposure time.

\section{Conclusions}

We presented far-infrared observations of the Lockman hole with AKARI satellite. 
In the performance verification phase, we performed slow scan observations 
in a 0.6 deg $\times$ 1.2 deg contiguous area overlapping with 
the ISO survey field and obtained number counts of galaxies down to 77, 26 
and 194 mJy (3$\sigma$) at 65, 90 and 140 $\rm{\mu m}$, respectively. 
The counts at 90 $\rm{\mu m}$ are $\sim$2 times deeper than previous 
measurements with ISO. Our results show that some of the ISO results 
should be reconsidered as described in Appendix. 

Our 90 $\rm{\mu m}$ counts were several times lower than the predicted counts 
by recent evolutionary models that show good agreement with the Spitzer/MIPS data. 
On the contrary, the observed counts at 65 and 140 $\rm{\mu m}$ are 
consistent with previous measurements with Spitzer at 70 and 160 $\rm{\mu m}$. 
Our 90 $\rm{\mu m}$ counts above 26 mJy accounts for $\sim$6\% of the CIB, 
and the bulk galaxies of the CIB may cause a steep rise of the counts below 
the detection limit. These results provide strong constraints on the 
evolutionary models.

The source counts presented in this paper were not color-corrected, because limited 
samples were available to examime the infrared colors of galaxies. Further study of 
the source counts could be done by proper color correction based on the measured 
SED of each source. A far-infrared deep survey in the lowest cirrus region with 
an area of $\sim$10 square degrees near the south ecliptic pole, as a part of 
AKARI deep survey programs (Matsuhara et al. 2006), is expected to provide us 
a large number of samples over 1000 sources down to $\sim$20 mJy at 90 $\rm{\mu m}$. 
Moreover, the AKARI all-sky survey will provide us extremely high statistics of 
galaxy samples on precedented sensitivity levels over the whole sky. 
The data sets obtained by such large-area surveys are promising not only for 
producing the SED templates for various galaxy populations but also for extending 
the flux range of the source counts to much brighter levels than we obtained 
in this work.

\bigskip
The AKARI mission is operated by Japan Aerospace Exploration Agency (JAXA), 
Nagoya University, the University of Tokyo and National Astronomical Observatory Japan, 
European Space Agency (ESA), Imperial College London, University of Sussex, 
Open University (UK), University of Groningen / SRON (Netherlands), and Seoul 
National University (Korea). The far-infrared detectors were developed under collaboration 
with The National Institute of Information and Communications Technology (Japan). 
The authors would like thank all the AKARI project members for their intensive effort.

\appendix
\section{Comparison with the ISO source catalogs in the Lockman hole}

\begin{figure}
  \begin{center}
    \FigureFile(120mm,120mm){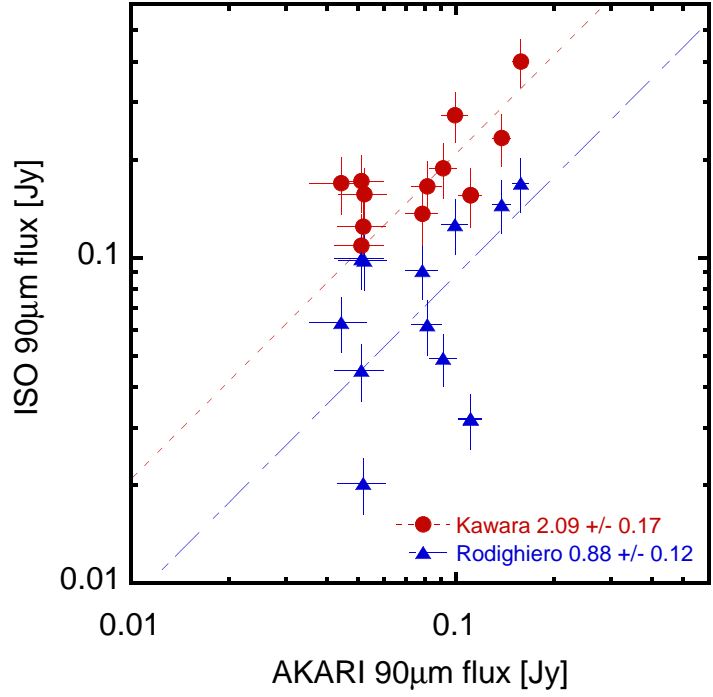}
  \end{center}
  \caption{Comparison of the AKARI measured flux for bright sources 
  in the Lockman hole with ISO results (Kawara et al. 2004, Rodighiero et al. 2005).}
  \label{fig:fig7}
\end{figure}

To confirm the flux calibration at lower flux levels, 
we compared the measured 90 $\rm{\mu m}$ fluxes of relatively 
bright sources with the ISO data for the same sources in Fig.~\ref{fig:fig7}. 
The dot-dashed and dashed lines are the best-fit lines with linear coefficients 
(scaling factor) of 0.88$\pm$0.12 for Rodighiero et al. (2005) and 2.09$\pm$0.17 
for Kawara et al. (2004), respectively. In terms of the absolute flux scale, 
the observed flux in Rodighiero et al. (2005) is consistent with 
ours within the calibration accuracy. However, relative deviation from 
the linear relation for Kawara et al. (2004), 0.17/2.09 = 0.08, is smaller 
than that for Rodighiero et al. (2005), 0.12/0.88 = 0.14, i.e., their relative 
calibration in the bright flux regime shows good agreement with ours.

A large difference between Kawara et al.'s count data and 
ours may arise from the fact that their absolute calibration relies 
on a single reference, IRAS F10507+5723, whose flux in the IRAS 
Faint Source Catalog (FSC) at 100 $\rm{\mu m}$ (1.22 Jy) is perhaps overestimated. 
In fact, the flux of this source measured with AKARI is 0.56$\pm$0.10 Jy 
at 90 $\rm{\mu m}$ and lower than the IRAS flux by a factor of $\sim$2. 
This factor is in good agreement with the scaling factor between 
Kawara et al. (2004) and our measurements for the sources as described above.
A tendency for the IRAS FSC to overestimate the flux is also pointed out 
by Heraudeau et al. (2004) in the data analysis for the ELAIS survey. 
Our recent result of the flux comparison between the AKARI all-sky survey data 
and the IRAS point source catalog (PSC) shows large scattering of 
the AKARI/IRAS flux ratio up to $\sim$5 even for bright sources 
at fluxes greater than 10 Jy in the IRAS PSC (Jeong et al. 2007). This result 
suggests the diffuculty of flux calibration with small number of samples of 
IRAS sources.

\begin{figure}
  \begin{center}
    \FigureFile(120mm,120mm){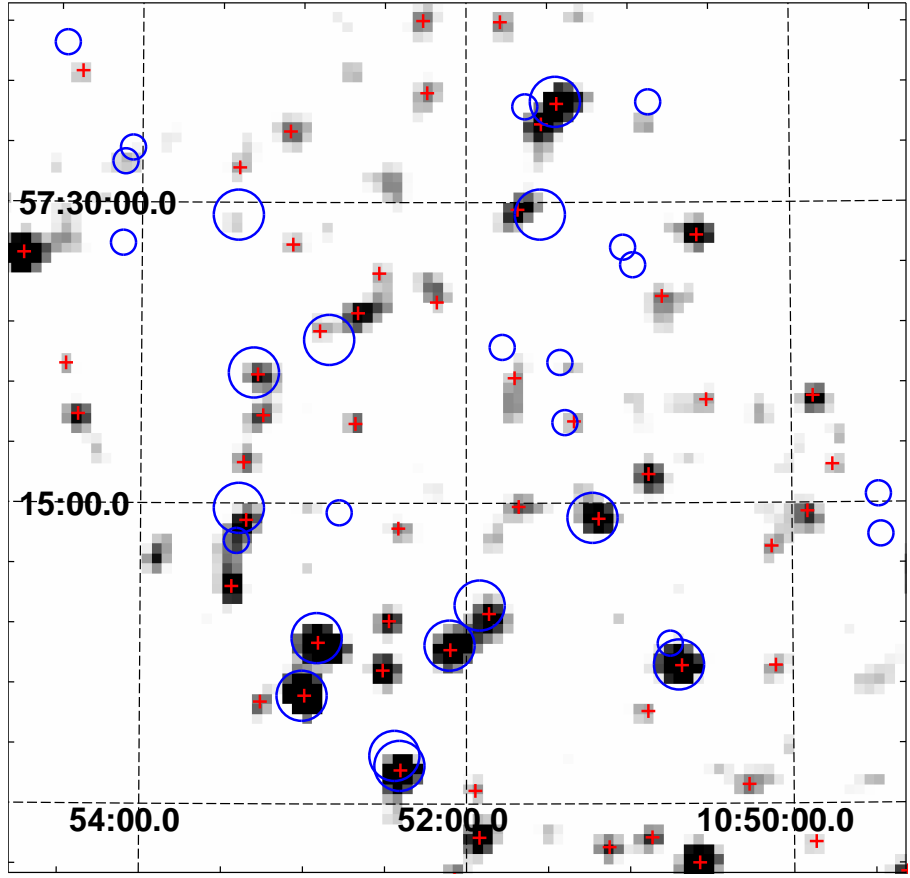}
  \end{center}
  \caption{Comparison of the source positions between Rodighiero et al. (2005) 
  and this work (background contour). In this linear contour map, darker pixels 
  have higher surface brightness as gray scale. Large circles denote brightest sources 
  listed in the catalog of Rodighiero et al. (2005) and also in Oyabu et al. (2005). 
  Small circles denote the remained faint sources in their catalog. Crosses are AKARI sources 
  detected at $S/N >$3. Many of the faint ISO sources are not identified by AKARI 
  and vice versa. In Fig.~\ref{fig:fig7}, all the ISO sources identified by AKARI are plotted.}
  \label{fig:fig8}
\end{figure}

A large deviation from the linear relation between the measured flux 
in Rodighiero et al. (2005) and ours seems to be related to the fact that 
some sources in their catalog are missing in our map. In Fig.~\ref{fig:fig8}, 
sources listed in their paper are plotted on the AKARI 90 $\rm{\mu m}$ map 
as background. Many of the sources in their catalog are not identified 
by AKARI and vice versa, though only bright sources are identified. 
Some of their sources were misidentified as faint sources detected 
by AKARI at $S/N <$3 and this fact results in large deviations in 
Fig.~\ref{fig:fig7}. Accordingly, there is no reason that their counts 
at fainter flux levels agrees with ours.

\clearpage

\begin{table}
  \caption{Parameters for aperture photometry}\label{tab:table1}
  \begin{center}
    \begin{tabular}{llllll}
      \hline
      Wavelength & Pixel size & Aperture radius & Net integration time & Point-source noise & Pixel noise \\
      $\rm{[\mu m]}$ & $\rm{[arcsec]}$ & $\rm{[arcsec]}$ & per pix $\rm{[s]}$ & (1$\sigma$) $\rm{[mJy]}$ & (1$\sigma$) $\rm{[MJy/sr]}$ \\ \hline
      65 & 30 & 37 & 12 & 26 & 0.20 \\
      90 & 30 & 39 & 12 &  9 & 0.06 \\
      140 & 50 & 58 & 20 & 76 & 0.39 \\
      160 & 50 & 61 & 20 & 288 & 0.60 \\
      \hline
    \end{tabular}
  \end{center}
\end{table}

\begin{table}
  \caption{Integral and differential counts at 90 $\rm{\mu m}$}\label{tab:table2}
  \begin{center}
    \begin{tabular}{lllllll}
      \multicolumn{3}{c}{Integral counts} & & \multicolumn{3}{c}{Differential counts} \\
            \cline{1-3}\cline{5-7}
      $S$ & Number & $N(>S)$ & & Average $S$ & bin width & $dN/dS \times S^{2.5}$ \\
      $\rm{[Jy]}$ &  & $\rm{[sr^{-1}]}$ & & $\rm{[Jy]}$ & $\rm{[Jy]}$ & $\rm{[sr^{-1}}$ $\rm{Jy^{1.5}]}$ \\
            \cline{1-3}\cline{5-7}
      0.119 &  4 & 1.92~e4 & & 0.140 & 0.042 & 3.39~e3 $\pm$ 1.70~e3 \\
      0.088 &  9 & 4.33~e4 & & 0.104 & 0.031 & 2.70~e3 $\pm$ 1.21~e3 \\
      0.065 & 18 & 8.66~e4 & & 0.077 & 0.023 & 3.10~e3 $\pm$ 1.03~e3 \\
      0.048 & 30 & 1.44~e5 & & 0.057 & 0.017 & 2.63~e3 $\pm$ 0.76~e3 \\
      0.036 & 45 & 2.16~e5 & & 0.042 & 0.013 & 2.11~e3 $\pm$ 0.54~e3 \\
      0.027 & 85 & 4.09~e5 & & 0.031 & 0.009 & 3.58~e3 $\pm$ 0.57~e3 \\
      \cline{1-3}\cline{5-7}
    \end{tabular}
  \end{center}
\end{table}

%
%
%
%
%


\end{document}